\documentclass[12pt,a4paper]{article}
\usepackage{amsmath,amsthm,amsfonts,amssymb,bbm}
\usepackage{graphicx,psfrag,subfigure}
\usepackage{cite}
\usepackage{hyperref}

\usepackage{color}

\numberwithin{equation}{section}
\numberwithin{figure}{section}

\newcommand{\Ai}{\mathrm{Ai}}
\newcommand{\Bi}{\mathrm{Bi}}
\newcommand{\Pb}{\mathbbm{P}}

\newcommand{\Id}{\mathbbm{1}}
\newcommand{\e}{\varepsilon}
\newcommand{\I}{{\rm i}}

\newcommand{\R}{\mathbb{R}}

\newcommand{\dx}{\mathrm{d}}
\newcommand{\A}{{\cal A}}
\newcommand{\Aflat}{{\cal A}_{\rm 1}}
\newcommand{\Acurved}{{\cal A}_{\rm 2}}

\newtheorem{prop}{Proposition}[section]
\newtheorem{thm}[prop]{Theorem}

\newtheorem{cla}[prop]{Claim}

\newtheorem{rem}[prop]{Remark}

\newenvironment{proofOF}[2]{\removelastskip\vspace{6pt}\noindent {\it Proof of #1.}~\rm#2}{\qed \par\vspace{6pt}}

\title{On the spatial persistence for Airy processes}
\author{Patrik L.\ Ferrari\thanks{Institute for Applied Mathematics, Bonn University, Endenicher Allee 60, 53115 Bonn, Germany. E-mail: {\tt ferrari@uni-bonn.de}} \and
Ren\'e Frings\thanks{Institute for Applied Mathematics, Bonn University, Endenicher Allee 60, 53115 Bonn,
Germany. E-mail: {\tt frings@uni-bonn.de}}
}

\date{October 17, 2012}

\begin{document}
\maketitle \sloppy

\begin{abstract}
In this short paper we derive a formula for the spatial persistence probability of the Airy$_1$ and the Airy$_2$ processes. We then determine numerically a persistence coefficient for the Airy$_1$ process and its dependence on the threshold.
\end{abstract}

\section{Introduction}\label{sectIntroduction}
The Airy$_1$ and Airy$_2$ processes are universal processes describing the fluctuation of interfaces for stochastic growth models in the Kardar-Parisi-Zhang (KPZ) universality class. The persistence probability is the probability that a process stays positive (resp.\ negative), or more generally, above (resp.\ below) a certain threshold during a time interval $[0,L]$. When the process is stationary, it might be expected that the persistence probability decays exponentially in $L$.

The Airy processes were obtained by studying specific models in the KPZ universality class~\cite{PS02,Jo03b,Sas05,BFPS06}. It was only in 2010 that in an amazing experiment with turbulent nematic liquid crystals Takeuchi and Sano~\cite{TS10,TSSS11} were able to verify experimentally the KPZ predictions at the level of distribution functions and covariances (and not only at the level of the scaling exponents). The agreement with the theory is very good.

In a recent paper the same authors~\cite{TS12} measured, among others, the spatial persistence coefficients with respect to a threshold given by the average of the process. In the case of the Airy$_2$ process, the persistence coefficients have been also measured in an off-lattice Eden model~\cite{Tak12} and verified by a numerical simulation of GUE Dyson's Brownian Motion~\cite{TS12}.

In this short paper we determine analytic formulas for the persistence probability to stay below a threshold $c$, both for the Airy$_1$ and the Airy$_2$ processes. The starting point are the two works on the continuum statistics~\cite{QR12,CQR11}. Then we focus on the case of the Airy$_1$ process and determine the associated persistence coefficient and its dependence on the threshold $c$. This is made by using the numerical approach for computing Fredholm determinants developed by Bornemann in~\cite{Born08}. The advantage of looking directly at the limit process is that we do not have uncontrolled uncertainties coming from the finite size settings of an experimental setup or of a numerical simulation.

\subsubsection*{Acknowledgments}
The authors would like to thank K.~Takeuchi for early discussions on his results and F.~Bornemann for giving advices on how to use his Matlab program. This work was supported by the German Research Foundation via the SFB611--A12 project.

\section{Results}\label{sectResults}
In order to state the results, let us introduce some notations. We denote by $\Aflat$ the Airy$_1$ process and by $\Acurved$ the Airy$_2$ process, see the review~\cite{Fer07} for the definition of these processes. For a threshold $c \in\R$ and a time interval $[0,L]$ with $L> 0$, the persistence probabilities are defined by
\begin{equation}
\begin{aligned}
P_-(\A,c,L) &= \Pb(\A(t)\leq c, 0\leq t\leq L),\\
P_+(\A,c,L) &= \Pb(\A(t)\geq c, 0\leq t\leq L),
\end{aligned}
\end{equation}
where $\A \in \{\Aflat,\Acurved\}$.

For large $L$, the persistence probabilities decay exponential in $L$ with persistence coefficients $\kappa_\pm$ given by
\begin{equation}\label{eqinterp}
P_\pm(\A,c,L)\simeq C_\pm(\A,c) e^{-\kappa_\pm(\A,c) L}\quad \textrm{for large }L.
\end{equation}
As it can be seen from Figure~\ref{FigAiry1Average} and Figure~\ref{FigAiry1Zero} below, the exponential decay of the persistence probabilities for the Airy$_1$ process is already observed at relatively small values of $L$, for instance already at $L=1$.

The analytic result for the persistence probabilities $P_-$ of the Airy processes are the following.
\begin{prop}\label{PropFlatKernel}
For the Airy$_1$ process we have
\begin{equation}\label{eqA1}
P_-(\Aflat,c,L)=\det(\Id-K_{1,L})_{L^2(\R)}
\end{equation}
where the kernel $K_{1,L}$ is given by
\begin{equation}\label{eq3.6}
K_{1,L}(x,y)=\Ai(|x|+y+2 c)+\Id_{[x\leq 0]} (\tilde K_{1,L}(x,y+2c)-\tilde K_{1,L}(-x,y+2c))
\end{equation}
with
\begin{equation}
\widetilde K_{1,L}(x,y)=\frac{1}{\sqrt{4\pi L}} \int_{\R_+} \dx z\, e^{-(x-z)^2/4L} e^{-2L^3/3} e^{-L(y+z)} \Ai(y+z+L^2).
\end{equation}
\end{prop}

\begin{prop}\label{PropCurvedKernel}
For the Airy$_2$ process we have
\begin{equation}
P_-(\Acurved,c,L)=\det(\Id-K_{2,L})_{L^2(\R)}
\end{equation}
where the kernel $K_{2,L}$ is given by
\begin{equation}
\begin{aligned}
K_{2,L}(x,y)&=K_{\rm Ai}(x+c,y+c)\\
&-\Id_{[x\leq 0]} \int_{\R_-}\dx z\, \int_{\R}\dx \mu\, e^{(\mu-c)L}
\phi(x,\mu)\phi(z,\mu) K_{{\rm Ai},L}(z+c,y+c)
\end{aligned}
\end{equation}
with
\begin{equation}
K_{{\rm Ai},L}(x,y)=(e^{L H_{\rm Ai}} K_{\rm Ai})(x,y)=\int_{\R_+}\dx \lambda\, e^{-L \lambda}\Ai(\lambda+x)\Ai(\lambda+y)
\end{equation}
and
\begin{equation}
\phi(x,\mu)=\frac{\Ai(\mu)\Bi(x+\mu)-\Ai(x+\mu)\Bi(\mu)}{\sqrt{\Ai(\mu)^2+\Bi(\mu)^2}}.
\end{equation}
\end{prop}

Before stating the results of the numerical evaluation of (\ref{eqA1}), let us resume the results cited above in the following table\footnote{The values for the Airy$_1$ process have to be multiplied by $2^{2/3}$ because the scaling in~\cite{TS12} is such that the limit process is
$u\mapsto 2^{2/3} \Aflat(2^{-2/3}u)$ instead of $u\mapsto \Aflat(u)$.}:
\begin{center}
\begin{tabular}{|l|c|c|c|}
  \hline
 &  $\A,c$ & $\kappa_-(\A,c)$ & $\kappa_+(\A,c)$\\
  \hline
Experimental~\cite{TS12} &  $\Aflat,-0.6033$ & $3.2(5)$ & $3.0(5)$ \\
Experimental~\cite{TS12} &  $\Acurved,-1.7711$ & $0.87(6)$ & $1.07(8)$ \\
Off-lattice Eden~\cite{Tak12} & $\Acurved,-1.7711$ & $0.89(4)$ & $0.90(2)$ \\
GUE Dyson's Brownian Motion~\cite{TS12} & $\Acurved,-1.7711$ & $0.90(6)$ & $0.90(8)$ \\
  \hline
\end{tabular}
\end{center}

While the different experiments and numerical simulations for the Airy$_2$ process provide results that are quite close to each other, no further results were available for the Airy$_1$ process. We first evaluated numerically the Fredholm determinant for the two natural thresholds, namely the average of the process\footnote{The Airy$_1$ process is a stationary process with one-point distribution given by $\Pb(\Aflat(0)\leq s)=F_1(2s)$ where $F_1$ is the GOE Tracy-Widom distribution function~\cite{FS05b}, and $F_1$ has an average $-1.20653$~\cite{TW96}.}, $c=-0.6033$, and for $c=0$, with the results
\begin{center}
\begin{tabular}{|c|c|c|c|}
  \hline
  $\A,c$ & $\kappa_-(\A,c)$ & $C_-(\A,c)$ \\
  \hline
  $\Aflat,-0.6033$ & $2.91$ & $0.370$ \\
  $\Aflat,\phantom{-}0\phantom{.0000}$ & $1.10$ & $0.733$ \\
  \hline
\end{tabular}
\end{center}
Comparing our result to the experimental one, we see that the agreement is fairly good. Indeed, the relative error for the Airy$_1$ process is 10\%.

\begin{figure}
\begin{center}
\psfrag{c}{$c$}
\psfrag{K}[c]{$\kappa_-(\Aflat,c)$}
\includegraphics[height=8cm]{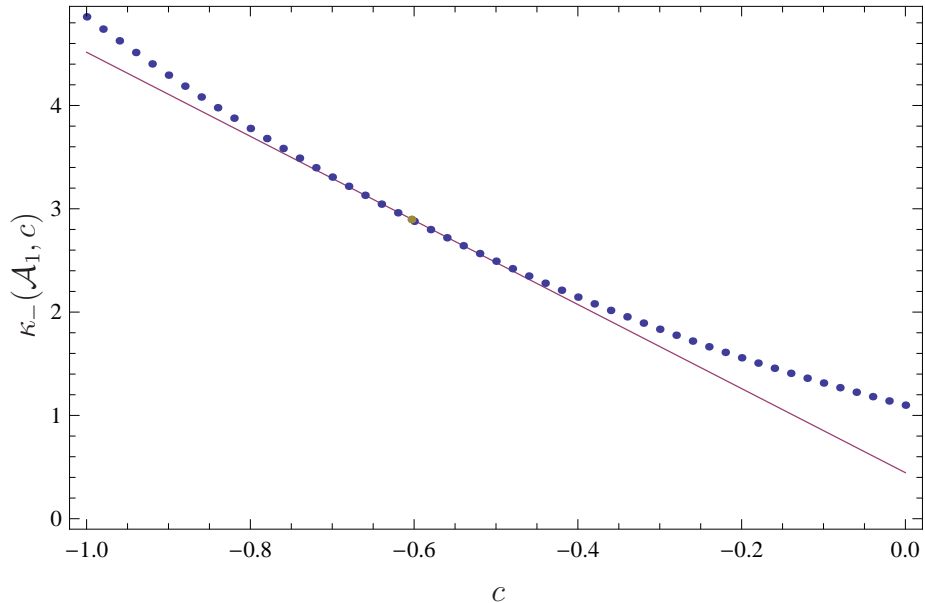}
\caption{Dependence of the $\kappa_-(\Aflat,c)$ as a function of $c$. The straight line is the linear approximation around the average of the Airy$_1$ process ($c=-0.6033$). The slope is $-4.07$.}
\label{FigkappaValues}
\end{center}
\end{figure}

We also determined the exact values of $\kappa_-(\Aflat,c)$ as a function of $c$ for $c\in [-1,0]$, see Figure~\ref{FigkappaValues}.
We observe that this coefficient is quite sensitive to the threshold $c$ (see Table~\ref{Table} for the values). For instance, in the region around the average of the process the tangent line has a slope of $-4.07$, i.e., a small error in the centering on the threshold leads to an error in the persistence coefficient $4$ times larger. According to the experimental results~\cite{TS10} and the analysis of specific models~\cite{FF11}, the approach to large time limit is at first order a constant times $t^{-1/3}$. Therefore, one has to take this finite size correction in account when setting the threshold (as it was made in~\cite{TS12}).

\section{Proofs of the analytic results}\label{sectAnalytic}
The starting point of our analysis are two formulas on the continuum statistics for the Airy$_1$ process~\cite{QR12} and for the Airy$_2$ process~\cite{CQR11}. Let us start with the Airy$_1$ process.
\begin{thm}[Theorem 4 of~\cite{QR12}]\label{ThmStartAiry1}
It holds
\begin{equation}\label{eqFlat}
\Pb(\Aflat(t)\leq g(t), 0\leq t\leq L)=\det(\Id-B_0+ \Lambda_{L,g} e^{-L\Delta} B_0)_{L^2(\R)}
\end{equation}
where $g$ is a function in $H^1([0,L])$, $\Delta$ is the Laplacian, $B_0(x,y)=\Ai(x+y)$, and
\begin{equation}
\Lambda_{L,g}(x,y)=\frac{e^{-(x-y)^2/(4L)}}{\sqrt{4 \pi L}}\Pb_{b(0)=x,b(L)=y}(b(s)\leq g(s), 0\leq s\leq L)
\end{equation}
with $b$ a Brownian Bridge from $x$ at time $0$ to $y$ at time $L$ and with diffusion coefficient $2$.
\end{thm}

To get the persistence probabilities, we have to determine the explicit kernel for the function $g(s)=c$.
\begin{proofOF}{Proposition~\ref{PropFlatKernel}}
We have to determine a formula for the Fredholm determinant of $\Id-B_0+\Lambda_{L,c} e^{-L\Delta} B_0$. Since the Fredholm determinant is on all $\R$, we can shift the variables by $c$ and obtain the kernel
\begin{equation}\label{eq3.9}
B_0(x+c,y+c)-\int_{\R} \dx z\, \Lambda_{L,c}(x+c,y+c) (e^{-L\Delta}B_0)(z+c,y+c).
\end{equation}
Clearly, $\Lambda_{L,c}(x,y)=\Lambda_{L,0}(x-c,y-c)$, therefore
\begin{equation}\label{eq3.10}
(\ref{eq3.9})=\Ai(x+y+2c)-\int_{\R} \Lambda_{L,0}(x,z) (e^{-L\Delta}B_0)(z+c,y+c).
\end{equation}
By the reflection principle we have
\begin{equation}
\begin{aligned}
\Lambda_{L,0}(x,z)&=\Pb_{b(0)=x,b(L)=z}(b(s)\leq 0, 0\leq s\leq L)\\
&= \frac{1}{\sqrt{4\pi L}}\left(e^{-(x-z)^2/(4L)}-e^{-(x+z)^2/(4L)}\right)\Id_{[x,z<0]}.
\end{aligned}
\end{equation}
Moreover, it is known (see e.g.\ the review~\cite{Fer07}) that
\begin{equation}
e^{-L \Delta} B_0(z+c,y+c)=e^{-2L^3/3-(z+y+2c)L}\Ai(z+y+2c+L^2).
\end{equation}
Putting all together we have
\begin{equation}\label{eq3.12}
(\ref{eq3.10})=\Ai(x+y+2c)-\Id_{[x<0]} \left(\widehat K_{1,L}(x,y+2c)-\widehat K_{1,L}(-x,y+2c)\right)
\end{equation}
where
\begin{equation}
\widehat K_{1,L}(x,y)=\frac{1}{\sqrt{4\pi L}} \int_{\R_-} \dx z\, e^{-(x-z)^2/4L} e^{-2L^3/3} e^{-L(y+z)} \Ai(y+z+L^2).
\end{equation}
Finally, using the identity (see below)
\begin{equation}\label{eq1}
\frac{1}{\sqrt{4\pi L}} \int_{\R} e^{-(x-z)^2/4L} e^{-2L^3/3} e^{-L(y+z)} \Ai(y+z+L^2) dz = \Ai(x+y)
\end{equation}
we get
\begin{equation}
\widehat K_{1,L}(x,y)=\Ai(x+y)-\widetilde K_{1,L}(x,y).
\end{equation}
Replacing this into (\ref{eq3.12}) gives the desired result (\ref{eq3.6}).

Finally, let us verify (\ref{eq1}). By the integral representation of the Airy function,
\begin{equation}\label{eqAiry}
\Ai(b^2+c)e^{2 b^3/3+bc} = \frac1{2\pi \I} \int_{e^{-\I\pi/3}\infty}^{e^{\I\pi/3}\infty}\dx w\, e^{w^3/3+b w^2-cw},
\end{equation}
for any $\e>0$, and a Gaussian integration we get
\begin{multline}
\frac{1}{\sqrt{4\pi L}} \int_{\R}\dx z\, e^{-(x-z)^2/4L} e^{-2L^3/3} e^{-L(y+z)} \Ai(y+z+L^2)\\
= e^{-L(x+y)} e^{L^3/3} \frac{1}{2\pi \I}\int_{e^{-\I\pi/3}\infty}^{e^{\I\pi/3}\infty} \dx w\, e^{w^3/3+L w^2-w(x+y-L^2)}=\Ai(x+y),
\end{multline}
where we used again (\ref{eqAiry}).
\end{proofOF}

Now we consider the Airy$_2$ process. The analogue of Theorem~\ref{ThmStartAiry1} for the Airy$_1$ process is given by
\newpage
\begin{thm}[Theorem 2 of~\cite{CQR11}]\label{ThmStartAiry2}
It holds
\begin{equation}\label{eqCurved}
\Pb(\Acurved(t)\leq g(t), 0\leq t\leq L)=\det(\Id-K_{\rm Ai}+ \Lambda_{L,g} e^{L H_{\rm Ai}}K_{\rm Ai})_{L^2(\R)}
\end{equation}
where $g$ is a function in $H^1([0,L])$, $H_{\rm Ai}=-\Delta+x$ is the Airy operator, $K_{\rm Ai}(x,y)=\int_{\R_+}\dx \lambda \, \Ai(x+\lambda)\Ai(y+\lambda)$ is the Airy kernel, and
\begin{multline}
\Lambda_{L,g}(x,y)\\
=e^{-Ly-L^3/3}\frac{e^{-(x-y)^2/(4L)}}{\sqrt{4 \pi L}}\Pb_{b(0)=x,b(L)=y-L^2}(b(s)\leq g(s)-s^2, 0\leq s\leq L)
\end{multline}
with $b$ a Brownian Bridge from $x$ at time $0$ to $y-L^2$ at time $L$ and with diffusion coefficient $2$.
\end{thm}
We have to determine the kernel for the special function $g(s)=c$.
\begin{proofOF}{Proposition~\ref{PropCurvedKernel}}
We have to compute the Fredholm determinant of $\Id-K_{\rm Ai}+ \Lambda_{L,c} e^{-L H_{\rm Ai}}K_{\rm Ai}$ over $L^2(\R)$.
As in the proof of Proposition~\ref{PropFlatKernel}, we first do a shift in the variables by $c$ and obtain the kernel
\begin{equation}\label{eq3.22}
K_{\rm Ai}(x+c,y+c)-\int_{\R}\dx z\, \Lambda_{L,c}(x+c,z+c) (e^{L H_{\rm Ai}}K_{\rm Ai})(z+c,y+c)
\end{equation}
It is easy to verify that
\begin{equation}
\Lambda_{L,c}(x,y)=\Lambda_{L,0}(x-c,y-c) e^{-L c}.
\end{equation}
Therefore, the kernel becomes
\begin{equation}
(\ref{eq3.22})=K_{\rm Ai}(x+c,y+c)-e^{-L c}\int_{\R}\dx z\, \Lambda_{L,0}(x,z) (e^{L H_{\rm Ai}}K_{\rm Ai})(z+c,y+c).
\end{equation}
Thus, the desired formula follows if we can show that
\begin{equation}
\begin{aligned}\label{eq2}
\Lambda_{L,0}(x,z) &= e^{-Lz-L^3/3}\frac{e^{-(x-z)^2/(4L)}}{\sqrt{4 \pi L}}\Pb_{b(0)=x,b(L)=z-L^2}(b(s)\leq -s^2, 0\leq s\leq L)\\
& = \Id_{[x,z\leq 0]} \int_\R \dx \mu\, e^{\mu L} \phi(x,\mu)\phi(z,\mu).
\end{aligned}
\end{equation}
To this end we use another representation of the kernel $\Lambda_{L,0}$, that can also be found in \cite{CQR11} and that follows from \eqref{eq2} by applying the Girsanov theorem and the Feynman-Kac formula. According to this characterization, $\Lambda_{L,0}(x,z)=u(L;x,z)\Id_{[z<0]}$ is the solution at time $t=L$ of the boundary value problem
\begin{equation}
\begin{aligned}
\partial_t u + H_{\Ai} u & = 0 \quad \text{ for } x<0 \text{ and } t\in(0,L), \\
u(0;x,z) & = \delta_{x-z},\\
u(t;x,z) & = 0 \quad \text{ for } x\geq0.
\end{aligned}
\end{equation}
The solution of this problem can be found in~\cite[eq.~(40)]{Mar98},
\begin{equation}
u(t;x,z) = \Id_{[x<0]}\int_\R \dx \mu\, e^{\mu t} \phi(x,\mu)\phi(z,\mu).
\end{equation}
Note that in \cite{CQR11} the boundary value problem describes the action of the operator $\Lambda_{L,0}$ while our formulation considers the kernel of this operator.
\end{proofOF}

\section{Numerical approach and results}\label{sectNumerical}
To apply the numerical procedure of~\cite{Born08} we need to have an analytic kernel, but the kernel in Proposition~\ref{PropFlatKernel} is not analytic at $x=0$. This issue can be fixed by rewriting the Fredholm determinant as acting on $L^2(\R)$ into a Fredholm determinant acting on $L^2(\R_-)\oplus L^2(\R_+)$. In this way, instead of a scalar non-analytic kernel we get an analytic $2\times 2$ matrix kernel.

There are few other issues that we had to deal with by trying to compute numerically the Fredholm determinant with kernel (\ref{eq3.6}):
\begin{itemize}
 \item We need to introduce a cut-off $T$ and compute the Fredholm determinant on $L^2([-T,T])$. We controlled the value of $T$ so that by varying it, the result were not changing.
  \item One can see that the kernel (\ref{eq3.6}) is not bounded, but this is not a relevant problem because the conjugated kernel obtained by multiplying (\ref{eq3.6}) with $e^{L(y-x)}$ is bounded.
  \item The main problem is that, even after conjugation, there are regions with the magnitude of the kernel that grows like $e^{a L^3}$ for $a$ of order $1$, while the kernel for positive $x$ has oscillations of order $1$. Consequently, the numerical approach works only for $L$ relatively small because of the limitation due to machine precision. For the range of $c\in [-1,0]$, it works well at least until $L=2.5$ (for $c=0$ also $L=3.5$ is still fine).
\end{itemize}
Fortunately, for the Airy$_1$ process, the logarithm of the persistence probability becomes rapidly a straight line as can be seen in Figure~\ref{FigAiry1Average} and Figure~\ref{FigAiry1Zero} below. This allowed us to determine the persistence coefficient $\kappa_-$ for the Airy$_1$ process reliably.
\begin{figure}
\begin{center}
\psfrag{L}{$L$}
\psfrag{P}[c]{$P_-(\Aflat,\textrm{mean},L)$}
\includegraphics[height=8cm]{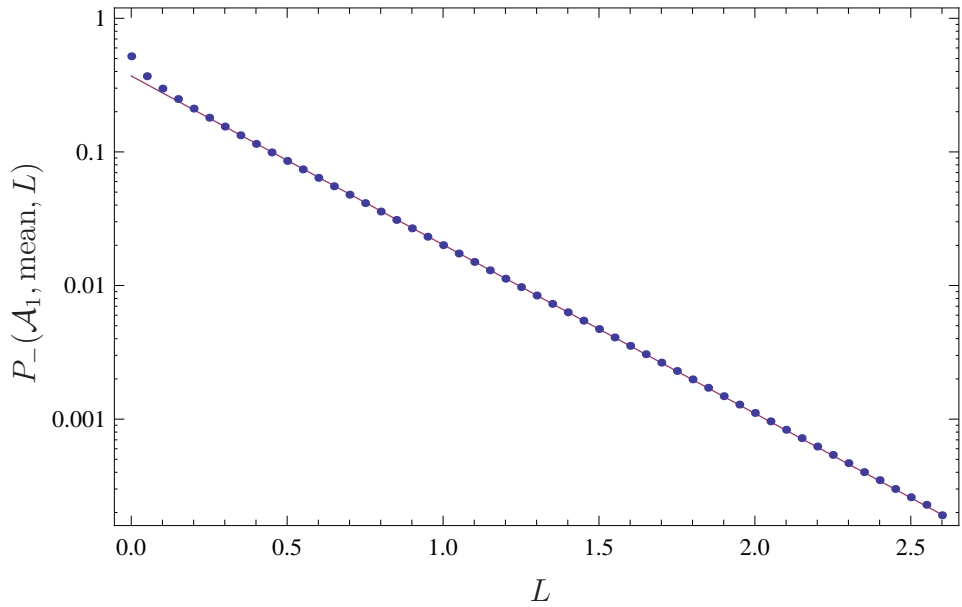}
\caption{Persistence probability for the Airy$_1$ process and exponential interpolation (\ref{eqinterp}) with $\kappa_-(-0.6033)=2.91$ and $C_-(-0.6033)=0.370$,}
\label{FigAiry1Average}
\end{center}
\end{figure}

\begin{figure}
\begin{center}
\psfrag{L}{$L$}
\psfrag{P}[c]{$P_-(\Aflat,0,L)$}
\includegraphics[height=8cm]{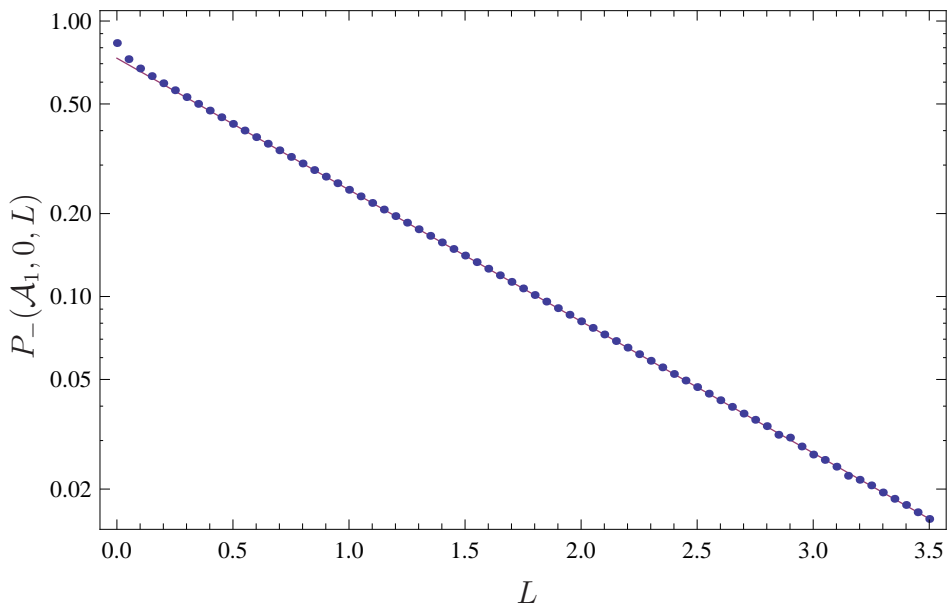}
\caption{Persistence probability for the Airy$_1$ and exponential interpolation  (\ref{eqinterp}) with $\kappa_-(0)=1.10$ and $C_-(0)=0.733$.}
\label{FigAiry1Zero}
\end{center}
\end{figure}

Finally, let us resume in a table the values of $\kappa_-(\Aflat,c)$ for $c\in[-1,0]$.
{\small \begin{table}[h]
\begin{tabular}{|c|c|c|c|c|c|c|c|c|c|c|}
  \hline
  $c$ & -1.00 & -0.98 & -0.96 & -0.94 & -0.92 & -0.90 & -0.88 & -0.86  & -0.84 & -0.82\\
  \hline
  $\kappa_-$ & 4.858 & 4.739 & 4.626 & 4.513 & 4.402 & 4.293 & 4.187 & 4.082 & 3.978 & 3.877\\
  \hline
  $c$  & -0.80 & -0.78 & -0.76 & -0.74 & -0.72 & -0.70 & -0.68 & -0.66 & -0.64 & -0.62\\
  \hline
  $\kappa_-$  & 3.778 & 3.680 & 3.584 & 3.490 & 3.398 & 3.307 & 3.218 & 3.131  & 3.045 & 2.961\\
  \hline
  $c$   & -0.60 & -0.58 & -0.56 & -0.54 & -0.52 & -0.50 & -0.48 & -0.46 & -0.44 & -0.42\\
  \hline
  $\kappa_-$  & 2.879 & 2.799 & 2.720 & 2.642  & 2.567 & 2.493 & 2.420 & 2.349 & 2.279 & 2.211\\
  \hline
  $c$   & -0.40 & -0.38 & -0.36 & -0.34 & -0.32 & -0.30 & -0.28 & -0.26 & -0.24 & -0.22\\
  \hline
  $\kappa_-$  & 2.145 & 2.080 & 2.016 & 1.954 & 1.893 & 1.834  & 1.776 & 1.719 & 1.664 & 1.610\\
  \hline
  $c$  & -0.20 & -0.18 & -0.16 & -0.14 & -0.12& -0.10 & -0.08 & -0.06 & -0.04 & -0.02 \\
  \hline
  $\kappa_-$  & 1.558 & 1.506 & 1.456 & 1.407 & 1.360 & 1.314 & 1.268 & 1.224 & 1.181 & 1.140 \\
  \hline
\end{tabular}
\caption{Values of $\kappa_-(\Aflat,c)$ for a set of values of $c\in[-1,0]$. The value of $\kappa_-(\Aflat,0)=1.099$.}
\label{Table}
\end{table}}


\begin{thebibliography}{10}

\bibitem{Mar98}
{Andres Martin-L\"of}, \emph{{The final size of a nearly critical epidemic, and
  the first passage time of a Wiener process to a parabolic barrier}}, J. Appl.
  Prob. \textbf{35} (1998), 671--682.

\bibitem{Born08}
F.~Bornemann, \emph{{On the numerical evaluation of Fredholm determinants}},
  Math. Comput. \textbf{79} (2009), 871--915.

\bibitem{BFPS06}
A.~Borodin, P.L. Ferrari, M.~Pr{\"a}hofer, and T.~Sasamoto, \emph{{Fluctuation
  properties of the TASEP with periodic initial configuration}}, J. Stat. Phys.
  \textbf{129} (2007), 1055--1080.

\bibitem{CQR11}
I.~Corwin, J.~Quastel, and D.~Remenik, \emph{{Continuum statistics of the
  Airy$_2$ process}}, arXiv:1106.2717 (2012).

\bibitem{Fer07}
P.L. Ferrari, \emph{{The universal Airy$_1$ and Airy$_2$ processes in the
  Totally Asymmetric Simple Exclusion Process}}, Integrable Systems and Random
  Matrices: In Honor of Percy Deift (J.~Baik, T.~Kriecherbauer, L-C. Li,
  K.~McLaughlin, and C.~Tomei, eds.), Contemporary Math., Amer. Math. Soc.,
  2008, pp.~321--332.

\bibitem{FF11}
P.L. Ferrari and R.~Frings, \emph{{Finite time corrections in KPZ growth
  models}}, J. Stat. Phys. \textbf{144} (2011), 1123--1150.

\bibitem{FS05b}
P.L. Ferrari and H.~Spohn, \emph{{A determinantal formula for the GOE
  Tracy-Widom distribution}}, J. Phys. A \textbf{38} (2005), L557--L561.

\bibitem{Jo03b}
K.~Johansson, \emph{Discrete polynuclear growth and determinantal processes},
  Comm. Math. Phys. \textbf{242} (2003), 277--329.

\bibitem{PS02}
M.~Pr{\"a}hofer and H.~Spohn, \emph{Scale invariance of the {PNG} droplet and
  the {A}iry process}, J. Stat. Phys. \textbf{108} (2002), 1071--1106.

\bibitem{QR12}
J.~Quastel and D.~Remenik, \emph{{Regularity and continuum statistics of the
  Airy$_1$ process}}, arXiv:1201.4709 (2012).

\bibitem{Sas05}
T.~Sasamoto, \emph{Spatial correlations of the {1D KPZ} surface on a flat
  substrate}, J. Phys. A \textbf{38} (2005), L549--L556.

\bibitem{Tak12}
K.A. Takeuchi, \emph{{Statistics of circular interface fluctuations in an
  off-lattice Eden model}}, J. Stat. Mech. (2012), P05007.

\bibitem{TS10}
K.A. Takeuchi and M.~Sano, \emph{{Growing interfaces of liquid crystal
  turbulence: universal scaling and fluctuations}}, Phys. Rev. Lett.
  \textbf{104} (2010), 230601.

\bibitem{TS12}
K.A. Takeuchi and M.~Sano, \emph{{Evidence for geometry-dependent universal
  fluctuations of the Kardar-Parisi-Zhang interfaces in liquid-crystal
  turbulence}}, J. Stat. Phys. \textbf{147} (2012), 853--890.

\bibitem{TSSS11}
K.A. Takeuchi, M.~Sano, T.~Sasamoto, and H.~Spohn, \emph{Growing interfaces
  uncover universal fluctuations behind scale invariance}, Sci. Rep. \textbf{1}
  (2011), 34.

\bibitem{TW96}
C.A. Tracy and H.~Widom, \emph{On orthogonal and symplectic matrix ensembles},
  Comm. Math. Phys. \textbf{177} (1996), 727--754.

\end{thebibliography}

\end{document}